\documentclass{iopart}
\usepackage{graphicx}
\newcommand{\+}[1]{#1^\dagger}
\newcommand{\xop}[1]{\hat #1^\times}
\newcommand{\oop}[1]{\hat #1^\circ}
\newcommand{\wn}{\ensuremath{\mathrm{cm}^{-1}}}

\begin{document}
\title{Correlated fluctuations in the exciton dynamics and spectroscopy of DNA}
\author{Arend~G.~Dijkstra and Yoshitaka Tanimura}
\address{Department of Chemistry, Graduate School of Science, Kyoto University, Kyoto 606-8502, Japan}
\eads{\mailto{dijkstra@kuchem.kyoto-u.ac.jp}, \mailto{tanimura@kuchem.kyoto-u.ac.jp}}
\begin{abstract}
 The absorption of ultraviolet light creates excitations in DNA, which subsequently start moving in the helix. Their fate is important for an understanding of photo damage, and is determined by the interplay of electronic couplings between bases and the structure of the DNA environment. We model the effect of dynamical fluctuations in the environment and study correlation, which is present when multiple base pairs interact with the same mode in the environment. We find that the correlations strongly affect the exciton dynamics, and show how they are observed in the decay of the anisotropy as a function of a coherence and a population time in a non-linear optical experiment.
\end{abstract}

\section{Introduction}

The bases that form the DNA molecule absorb ultraviolet light, leading to a surplus of energy that can cause damage. Fortunately for all living organisms that depend on this molecule, the excess energy is dissipated rapidly. The existence of delocalized states has been suggested to play an important role in the dissipation mechanism \cite{Fiebig.2009.jpcb.113.9348}.
Evidence for delocalized excitations was found using femtosecond time-resolved spectroscopy \cite{Onidas.2007.pccp.9.5143, Buchvarov.2007.pnas.104.4794}, as well as from combined density functional and molecular dynamics simulations \cite{Bouvier.2003.jpcb.107.13512, Tonzani.2008.jacs.130.7607}.
 
Spectroscopic studies reveal a shift of 5000-7000 \wn in the energy of an exciton in DNA after excitation. This shift is observed as a Stokes shift, the difference between the maxima of the absorption and steady state fluorescence spectra \cite{Markovitsi.2006.jppa.183.1}, as well as by a time-resolved shift in the excited state absorption \cite{Buchvarov.2007.pnas.104.4794}. In the latter experiment, the wavelength of the excited state absorption is found to shift from 380 to 330 nm, which, in the case of localized final states, is equal to the shift in the energy of the created excitation. 

This observed energy shift could be explained in a band model as follows.
Because the transition dipoles in stacked bases make only a small angle (36 degrees in B-DNA) with each other, and positive values have been reported for the couplings, the DNA stack forms an H-aggregate. In this case, absorption is mainly at the top of the excitonic band, while emission occurs from lower-lying states. This qualitatively explains the observed shift. Quantitatively, however, the maximum Stokes shift in such a model is given by the exciton band width. As a rough estimate, which does not include the long-range interactions, this is equal to four times the coupling strengths. Most reported values of the couplings are smaller than 300 \wn. Even when the long-range interactions are included, significantly larger couplings would be needed to explain the observed Stokes shift. More naturally, however, the shift can be explained from an interaction with the environment, which increases the average bandwidth.

Such an interaction has been included by considering the effect of a static disordered environment on the dynamics of excitons and charge transfer states \cite{Bittner.2006.jcp.125.094909, Bittner.2007.jppa.190.328}.
In the case of charge transport in DNA, which is often interpreted in band models \cite{Porath.2000.nature.403.635}, the effects of disorder have been studied extensively. The effect of a static environment on the excitation energies \cite{Malyshev.2009.jpcm.21.335105}, as well as static variations in the structure of the DNA \cite{Roche.2003.prl.91.108101} have been included in calculations of the conductance. Studies have highlighted the role of correlated fluctuations, present when multiple bases in DNA interact with the same mode in the environment, and their effect on the electronic properties \cite{Carpena.2002.nature.418.955, Diaz.2009.cp.365.24}.

In the static description, the environment does not change on the time scale of the experiment. Its state is, however, different for each helix in the ensemble probed in a measurement. In turn, the system properties such as the transition energies of each base, or the couplings between bases, vary from helix to helix. Their values for an individual system are often called realizations of the disorder, and the properties of the ensemble can be found by averaging over all possible realizations, weighted by their respective probabilities of occurrence. In the static situation, the eigenstates of the Hamiltonian are well-defined, and can be used as the first step in understanding the properties of the system.

However, in the case of molecules in solution, the environment is highly dynamic. The motion of the environment will lead to time dependence of the transition frequencies and couplings in the system (fluctuations), as well as to the exchange of energy with the environment (dissipation). The eigenstates of the system, found by diagonalizing the Hamiltonian for a given configuration of the bath, will only exist for a short time. As soon as the bath changes its state, the original eigenstates will mix. This situation can be called a dynamic bath. 

The effect of such a dynamic bath on the charge transport has been calculated from molecular dynamics simulations combined with quantum chemical methods \cite{Gutierrez.2009.prl.102.208102, Woiczikowski.2009.jcp.130.215104}. Dynamic disorder (in the form of conformational fluctuations) is found to be important, as models with only static disorder cannot realistically describe charge transfer in DNA. These studies also strenghten the expectation that correlations in the fluctuations might be important in DNA. In the case of excitons, the coupling to a dynamic environment has been shown to be reflected in non-linear optical observables \cite{Kim.2008.jcp.128.135102}.

The interplay of electronic coupling with fluctuations and dissipation, induced by an environment, is regarded as an important problem in chemical physics. The resulting energy transport is traditionally modelled with F\"orster theory, but this cannot describe the energy transfer in the intermediate coupling limit or the presence of correlated fluctuations \cite{Beljonne.2009.jpcb.113.6583}. In the context of quantum networks \cite{Cao.2009.jcpa.113.13825}, it has been realized that the presence of fluctuations can increase the efficiency of excitation transport \cite{Mohseni.2008.jcp.129.174106, Caruso.2009.jcp.131.105016}. Recently, using two-dimensional optical spectroscopy, it has been discovered that quantum coherence in photosynthetic light-harvesting complexes can live much longer than expected from perturbative treatments of the system-bath interaction \cite{Engel.2007.nature.446.782}. Although it was previously suggested that the origin of this effect might be found in correlated disorder, more recent studies have found that long-lived coherence can originate in a proper treatment of the time scale of bath fluctuations \cite{Ishizaki.2009.jcp.130.234111}. These treatments have, however, been limited to uncorrelated disorder. The effect of interplay between a dynamic bath and correlated fluctuations is yet unknown.

This highlights the importance of a proper understanding of the quantum dynamics of a system in contact with its environment, including the possibility of correlated fluctuations.
The effects of an environment that evolves on a time scale comparable to the excitonic dynamics, as well as the presence of environmental modes that couple to different chromophores simultaneously, remains to be studied. 
Rigorous theories have been developed to deal with the dynamics of a system in contact with a dynamic, quantum mechanical bath. The hierarchy of equations of motion approach \cite{Tanimura.1989.jpsj.58.101, Ishizaki.2005.jpsj.74.3131, Tanimura.2006.jpsj.75.082001, Ishizaki.2007.jpca.111.9269, Ishizaki.2009.jcp.130.234111,Strumpfer.2009.jcp.131.225101, Chen.2009.jcp.131.094502} can be used to include dynamical fluctuations, as well as dissipation.
Numerical path integral approaches that can deal with the dynamic effects induced by a quantum mechanical bath were introduced \cite{Mak.1991.pra.44.2352,  Topaler.1996.jpc.100.4430, Makri.1998.jpca.102.4414, Sim.2004.jpcb.108.19093}, and applied to the charge transfer in pieces of DNA \cite{Kim.2008.jpcb.112.2557}.

In this paper, we investigate the exciton dynamics in DNA in the presence of a dynamic bath. We focus on the effect of correlations in the fluctuations on the dynamics, and show how their effects are observable in the non-linear optical response. In particular, we calculate the anisotropy decay as a function of a coherence time and a population time, which has been used in experiment to characterize exciton transfer in conjugated polymers \cite{Collini.2009.science.323.369}.
We introduce the theory, which includes a proper treatment of the environmental time scale as well as the presence of correlated fluctuations in section \ref{sec:model}. In section \ref{sec:results} we present the resulting exciton dynamics and calculated linear absorption spectra and two-time anisotropy decay data. Section \ref{sec:concl} contains concluding remarks.

\section{Model} \label{sec:model}

\subsection{Hamiltonian and equations of motion}

\begin{figure}[t]
 \includegraphics[width=10cm]{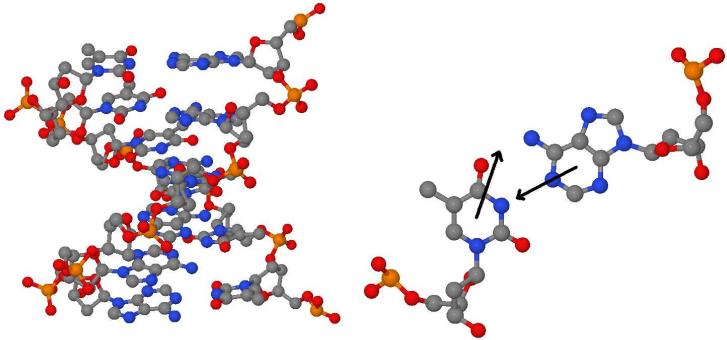}
\caption{\label{fig:dna} Structure of the six base pair DNA helix and of a single base pair, with the transition dipoles indicated. The DNA structure was generated with 3DNA \cite{x3dna}.}
\end{figure}

In this paper, we consider a piece of six base pairs of poly(dA) poly(dT) B-DNA, which has been studied experimentally and theoretically as a model compound. In this structure, shown in figure \ref{fig:dna}, each base pair contains a dA and a dT base, and all base pairs lie flat in planes perpendicular to the helix axis. The relative orientation of two adjacent base pairs is defined by the twist angle. Although this angle can fluctuate considerably in solution, our model assumes a fixed twist angle of 36 degrees, which is the average value for B-DNA. For this value, the helix makes a full turn every 10 base pairs.

In each base, we include a single strong ultraviolet transition, which is modelled as a two-level system. The frequencies and transition dipoles are chosen following \cite{Bouvier.2002.cp.275.75}. The transition in the dT base is between the $S_0$ and $S_1$ states, and has a frequency of $\epsilon_T = 37.500\,\wn$. In the dA base, the strongest absorption is found for the $S_0 \to S_2$ transition, with a frequency of $\epsilon_A = 38.800\,\wn$. Both transitions have a transition moment of $3.7 D$ in water. The transition dipoles lie in the plane of the bases, but their directions
are not well known. Experimental and calculated results vary over about 90 degrees for adenine, and 30 degrees for thymine. To obtain a definite model, we adopt the experimentally determined values, as shown in figure \ref{fig:dna}. 
In adenine, the dipole $\vec \mu_{nA}$ makes an angle of 66 degrees with the C-NH2 bond. In thymine, we use +14 degrees from the C=O bond for the dipole $\vec \mu_{nT}$. This results in an angle of 117 degrees between the two transition dipoles in each base pair.

The base pairs are labelled with indices $n$ and $m$. The standard exciton Hamiltonian, which contains the excitation energies of the $2 N$ bases, as well as interactions between them, is given in terms of the creation and annihilation operators $\+{c}$ and $c$ by
\begin{equation}
  H_S = \sum_{n=1}^N \sum_{i\in\{A,T\}} \epsilon_i \+{c}_{ni} c_{ni} + \sum_{n,m=1}^{N} \sum_{i,j\in\{A,T\}} J_{ni,mj} \+{c}_{ni} c_{mj}. 
\end{equation}
In the B-DNA structure, the couplings were estimated to be $J_{nA, nT} = 248\,\wn$, $J_{nA,(n+1)A} = 217\,\wn$ and $J_{nT,(n+1)T} = 170\,\wn$ \cite{Bouvier.2002.cp.275.75}. The couplings between stacked T bases and between the A and T in the same base pair agree well with results from quantum chemical calculations. Using these methods, however, a much larger value ($872\,\wn$) was found for the interaction between stacked A bases \cite{Czader.2008.jcp.128.035101}. This 5 times larger value would significantly increase the length over which coherence can be present in the stack of A bases. In addition, the effect of interactions between non-adjacent bases must be considered. Here, we will not try to model the coupling strengths in more detail, but restrict our discussion to the parameters given above.

In addition to the direct interactions between the base pairs, the excitations in DNA strongly interact with the environment, as can be seen from the Stokes shift observed in experiment \cite{Markovitsi.2006.jppa.183.1}. The environment is modelled as a collection of independent harmonic oscillators, labelled by an index $\alpha$, with masses $m_\alpha$, coordinates $x_\alpha$, momenta $p_\alpha$ and frequencies $\omega_\alpha$, which gives the bath Hamiltonian $H_B = \sum_\alpha \left( p_\alpha^2/2 m_\alpha + m_\alpha \omega_\alpha^2 x_\alpha^2/2 \right)$.
The interaction between the bases and the environment is given by
\begin{equation}
H_{SB} = - \sum_{n=1}^N \sum_{i\in\{A,T\}} \sum_\alpha g_{ni,\alpha} V_{ni} x_\alpha,
\end{equation}
where $V_{ni} = \+{c}_{ni} c_{ni}$.
This system-bath interaction leads to fluctuations in the excitation frequencies of individual bases, and to dissipation of energy into and out of the bath. It describes the dynamics of excitons in the system, including their relaxation within the excitonic bands. We do not include couplings of the bath to a single system creation or annihilation operator, which changes the number of excitons in the system, and eventually returns the system to the ground state.
The information on the system bath coupling is contained in the parameters $g_{ni,\alpha}$, which denote the effect of the bath mode with frequency $\omega_\alpha$ on the transition energy of the $ni^\mathrm{th}$ base. Because the bath modes are harmonic oscillators, all information on the coupling of the system to the bath is contained in the spectral densities
\begin{equation}
  \mathcal J_{ni,mj}(\omega) = \sum_\alpha \frac{g_{ni, \alpha} g_{mj,\alpha}}{2 m_\alpha \omega_\alpha} \delta(\omega - \omega_\alpha).
\end{equation}

A useful model for the spectral densities, which employs a single bath time scale for each base, is given by the overdamped Brownian oscillator,
\begin{equation}
  \mathcal J_{ni,mj}(\omega) = 2 \lambda_{ni,mj} \gamma_{ni,mj}  \frac{\omega \gamma_{ni,mj}}{\gamma_{ni,mj}^2 + \omega^2}.
\end{equation}
In principle, our treatment is not limited to overdamped modes, the approach used here can be extended to treat a more general Brownian spectral distribution \cite{Tanimura.1994.jpsj.63.66, Tanaka.2009.jpsj.78.073802}.
In the overdamped case, and assuming the high temperature limit, the correlation functions for the effective bath modes are given by \cite{Tanimura.1989.jpsj.58.101, Tanimura.2006.jpsj.75.082001}
\begin{equation}
 L_{ni,mj}(t) = c_{ni,mj} e^{- \gamma_{ni,mj}|t|},
\end{equation}
with
\begin{equation} \label{cfcoeff}
  c_{ni,mj} = \lambda_{ni,mj}(-i \gamma_{ni,mj} + \frac{2}{\beta}).
\end{equation}
The functions $L_{ni,ni}(t)$ are autocorrelation functions.  Their real parts describe the magnitude and the time scale of the fluctuations in the $ni^\mathrm{th}$ base energy, determined at a given temperature $T = 1/ k_B \beta$ by $\lambda_{ni,ni}$ and $\gamma_{ni,ni}$, respectively. The imaginary parts of the correlation functions are responsible for the dissipation of energy. The presence of correlations between fluctuations on different bases is modelled by the cross correlation functions $L_{ni,mj}$ for $ni \neq mj$. They determine the degree to which fluctuations in the $ni^\mathrm{th}$ and the $mj^\mathrm{th}$ frequency are correlated. If these quantities are all zero, all base energies fluctuate independently. In the other extreme case, the fluctuations are perfectly correlated if $\lambda_{ni,mj} = \lambda_{ni,ni} = \lambda_{mj,mj}$ and $\gamma_{ni,mj}=\gamma_{ni,ni}=\gamma_{mj,mj}$. In this case, the $ni^\mathrm{th}$ and $mj^\mathrm{th}$ base are coupled to a common bath, with the same strength.

The dynamics generated by the complete Hamiltonian, which we will call the Hamiltonian of the medium, $H_\mathrm{M} = H_\mathrm{S} + H_\mathrm{B} + H_\mathrm{SB}$, is given by the Liouville equation
\begin{equation} \label{liouville}
  \dot R(t) = -i \xop H_M R(t),
\end{equation}
where $\xop A B = [A, B]$ and $R(t)$ is the density matrix that contains all the degrees of freedom of the system as well as the bath. Note that at this point the equation of motion is entirely equivalent to the Schr\"odinger equation, where the same Hamiltonian $H_\mathrm{M}$ describes the time evolution of a wave function. The formal solution of (\ref{liouville}) is given in terms of the propagator $G(t; t_0) = \exp(-i \xop H_M (t-t_0))$ by $R(t) = G(t; t_0) R(t_0)$. This allows, in principal, the calculation of the density matrix $R(t)$ at all times. The interesting observables are, however, normally defined only in terms of the $2 N$ degrees of freedom of the system, and do not require the knowledge of the infinite-dimensional density matrix $R(t)$. One therefore introduces the reduced density matrix, which only includes the system's degrees of freedom, as $\rho(t) = \tr_B R(t) = \sum_\alpha \langle \alpha | R(t) | \alpha \rangle$. The time evolution of this reduced density matrix for a system linearly coupled to a harmonic bath can be found using numerical path integral techniques \cite{Topaler.1996.jpc.100.4430, Makri.1998.jpca.102.4414,Mak.1991.pra.44.2352}. An alternative and efficient method, which we will use here, employs a set of equations of motion for the system's reduced density matrix and multiple auxiliary density matrices \cite{Tanimura.1989.jpsj.58.101, Ishizaki.2005.jpsj.74.3131, Tanimura.2006.jpsj.75.082001, Ishizaki.2007.jpca.111.9269, Ishizaki.2009.jcp.130.234111,Strumpfer.2009.jcp.131.225101, Chen.2009.jcp.131.094502, Tanaka.2009.jpsj.78.073802}. It can be derived starting from the Feynman-Vernon influence functional approach \cite{Tanimura.1989.jpsj.58.101, Tanimura.2006.jpsj.75.082001}. The reduced density matrix for the electronic transitions is given by path integrals over coordinates $\phi(t)$ and $\phi'(t)$. The influence functional at time $t$, when the system was prepared in a factorized state at time $0$, can be calculated by generalizing the procedure described in \cite{Leggett.1987.rmp.59.1}, and is found to be
\begin{eqnarray}
  F[\{\phi(t)\}] &=& \sum_{ni, mj} \int_0^t \mathrm{d}t' \int_0^{t'} \mathrm{d}t'' \xop V_{ni}[\{\phi(t')\}] \left( \mathrm{Re} L_{ni,mj}(t'-t'') \xop V_{mj}[\{\phi(t'')\}] \right. \nonumber \\ 
  &+& \left.  i \mathrm{Im} L_{ni,mj}(t'-t'') \oop V_{mj}[\{\phi(t'')\}] \right),
\end{eqnarray}
where $\{\phi\}$ denotes the pair $\phi$ and $\phi'$, $V_{ni}[\phi(t)]$ is the representation of the operator $V_{ni}$, $\xop V_{ni}[\{\phi(t)\}] = V_{ni}[\phi(t)] - V_{ni}[\phi'(t)]$ and $\oop V_{ni}[\{\phi(t)\}] = V_{ni}[\phi(t)] + V_{ni}[\phi'(t)]$.

Assuming commuting operators $V_{ni}$, a hierarchy of equations of motion can be derived by introducing auxiliary density matrices, indexed by a set of indices $n_{ss'}$. The reduced density matrix for the electronic transitions is found by setting all the indices to zero. It nonperturbatively contains the effects of the bath, as ensured by the presence of the auxiliary density matrices.
In the high-temperature approximation \cite{Tanimura.1989.jpsj.58.101}, the hierarchy including cross-correlation terms, is given by
\begin{eqnarray}
 \dot \rho^{\{n\}}(t) 
   &=& - \left( i \xop H_S + \sum_{ss'} n_{ss'} \gamma_{ss'} \right) \rho^{\{n\}}(t) \nonumber \\
  &-& i \sum_{ss'} n_{ss'} \left( c_{ss'} V_{s'} \rho^{n_{ss'}^-}(t) - c_{ss'}^* \rho^{n_{ss'}^-}(t) V_{s'} \right) - i \sum_{ss'} \xop V_{s} \rho^{n_{ss'}^+}(t) \label{heom} ,
\end{eqnarray}
where we have introduced the notation $s$ and $s'$ for the pair $ni$ and $n_{ss'}^{\pm} = n_{ss'} \pm 1$. The coefficients $c_{ss'}$ are given by (\ref{cfcoeff}). For systems at lower temperature, the hierarchy can be extended with low temperature correction terms \cite{Ishizaki.2005.jpsj.74.3131, Tanimura.2006.jpsj.75.082001}. The lowest member of the hierarchy, which corresponds to the physical reduced density matrix, fully includes the coupling of the system to the bath. The deeper layers can be understood as bookkeeping devices which store the state of the bath, and, importantly, the coherences between system and bath states, at earlier times. 

To simplify the physical picture as well as the numerics, we will discuss only cases in which the fluctuations on two bases are either uncorrelated, or completely correlated. We will furthermore assume that the strength and the time scale of the fluctuations in each base are the same. In this case, the equations of motion simplify to the form
\begin{eqnarray}
 \dot \rho^{\{n\}}(t) 
   &=& - \left( i \xop H_S + \sum_{s} n_{s} \gamma \right) \rho^{\{n\}}(t) \nonumber \\
  &-& i \sum_{s} n_{s} \left( c V_{s} \rho^{n_{s}^-} - c^* \rho^{n_{s}^-} V_{s} \right) - i \sum_{s} \xop V_{s} \rho^{n_{s}^+} \label{heomuncorr},
\end{eqnarray}
with $c = \lambda(-i \gamma + \frac{2}{\beta})$. The sum over $s$ runs over the terms in the system bath coupling. If all fluctuations are uncorrelated, $s$ denotes a base
$ni$, and $V_s = \+{c}_{ni} c_{ni}$. The model then includes 12 system bath operators $V$ for a six base pair helix. In the case where the fluctuations in $\epsilon_A$ and $\epsilon_T$ on the same base pair are perfectly correlated, but uncorrelated with the frequencies on other base pairs, $s$ refers to a base pair $n$, and $V_s = \+{c}_{nA} c_{nA} + \+{c}_{nT} c_{nT}$. We furthermore approximately include the low temperature terms as described in \cite{Ishizaki.2009.pnas.106.17255}. The dynamics obtained from this equation of motion fully includes the fluctuations and dissipation induced by the interaction with the bath, without relying on a perturbative or fast bath approximation.

In the simulations presented in this paper, we use $\lambda = 2.5 \cdot 10^3\,\wn$ and a bath time scale $\tau_B = 1/ \gamma = 50$ fs. The choice of $\lambda$, which is an order of magnitude larger than the couplings $J$, is motivated by the observed large Stokes shift \cite{Markovitsi.2006.jppa.183.1, Buchvarov.2007.pnas.104.4794}. The temperature is set to 300 K. For these values, $\beta \hbar \gamma = 0.5$, which confirms the validity of the high-temperature approximation.
 
\subsection{Optical response}
\subsubsection{Linear response.}
The dynamics of a single excitation in the DNA can be obtained directly by propagating (\ref{heom}) from a given initial condition.
We now turn to the calculation of optical observables, which is formulated by combining the propagation the equation of motion with the correct sequence of matter field interactions. These are deduced by coupling the system to an external electric field $\vec E(t)$. The total Hamiltonian is the sum of the Hamiltonian $H_M$ (defined as $H_\mathrm{M} = H_\mathrm{S} + H_\mathrm{B} + H_\mathrm{SB}$), the Hamiltonian for the field and an interaction term, which is given in the semi-classical description and in the dipole approximation by
\begin{equation}
  H_\mathrm{ML} = - \mu \cdot \vec E(t) = -\sum_n \vec \mu_n \cdot \vec E(t) (\+{c}_n + c_n).
\end{equation}
Note that in this equation we have slightly condensed our notation and used indices $n$ and $m$ to denote a single base rather than a base pair. This convention will be used throughout this section.
The linear response requires the propagation of the free dynamics of the medium (i.e. in the absence of the field) after a single interaction with the light at a time $t_0$. The resulting polarization at time $t$ can be expressed as the convolution of the electric field with a response function \cite{Mukamel.1995.book}. Assuming that the excitation is created by interacting with an ultra short laser pulse, the convolution becomes trivial, and the measurement of the polarization directly gives the response function as $S(t) = \langle \tr \mu \rho(t) \rangle_O$. The notation $\langle \cdots \rangle_O$ indicates that the response function is averaged over all orientations of the sample with respect to the fixed coordinate system of the laboratory. The reduced density matrix is found from the perturbation expansion $\rho(t) = i \tr_B G(t; t_0) \xop \mu R(t_0)$ (for $t > t_0$). After making the rotating wave approximation (RWA), which is required for a consistent treatment of ultra short pulse excitation \cite{Mukamel.1995.book}, it reduces to
\begin{equation}
  \rho(t) = i \tr_B G(t; t_0) \mu R(t_0).
\end{equation}
As can be seen, this reduces the number of terms in the linear response from two ($\xop \mu R = \mu R  - R \mu$) to only one. 

Finally, the linear absorption spectrum is given as the Fourier transform of the response function (using $t_0 = 0$),
\begin{equation} \label{linspec}
  A(\omega) = \frac{1}{3} \mathrm{Re} \int_0^\infty \mathrm{d}t e^{i \omega t} \vec\mu_n \cdot \vec\mu_m G_{n0;m0}(t; 0).
\end{equation}
The factor $1/3$ originates in the average over orientations. The indices $n$ and $m$, which label individual bases, are understood to be summed over, and $0$ denotes the ground state. We have assumed that the system is in the ground state before interacting with the electric field. This is entirely reasonable in the case of optical excitation of DNA. The excitation energy is orders of magnitudes larger than the thermal energy, preventing thermal population of other states than the ground state. The interaction with the field then puts the system into a coherent superposition of a singly excited state and the ground state. Because the medium Hamiltonian conserves the number of excitations, only the time evolution of such coherent superpositions is required for the calculation of the linear response. This is indicated by the notation $G_{n0; m0}(t; 0)$, where $n$ and $m$ denote the excitation of single bases, and
\begin{equation}
  \langle n | \rho(t) | 0 \rangle = i \tr_B G_{n0; m0}(t; 0) \vec\mu_m R(t_0).
\end{equation}
It is essential to note that the propagator $G$ does not only include the degrees of freedom of the system. It is written in the product basis of the system degrees of freedom and all bath states, and explicitly mixes the system and the bath through the action of $H_{SB}$. All mixing processes are included nonperturbatively by propagating the hierarchy of equations of motion starting from $\vec\mu_m R(0)$, which boils down to the calculation of $G_{n0; m0}(t; 0)$.

\subsubsection{Third-order response.}
Because the linear absorption contains no direct information about the exciton dynamics, and the second-order response vanishes in an isotropic medium, we will furthermore employ the third-order response \cite{Mukamel.1995.book, Cho.2008.chemrev.108.1331}. Although it can be derived following the same procedure as outlined above for the linear response, the calculations is naturally somewhat more involved. More importantly, a third-order response function depends on three time intervals, and can be written generally as $S(t_1, t_2, t_3)$. A proper description of the system-bath interaction, which is not limited to an ultra fast bath, includes the memory stored in the bath of the system state at earlier times. The presence of this memory makes it impossible to split the reduced response functions, which depend only on the system's degrees of freedom, into functions that depend only on a single time variable, $S(t_1, t_2, t_3) \neq S(t_1) S(t_2) S(t_3)$. The memory in the bath strongly affects the lineshape in nonlinear observables. Although it is quite naturally included in the simulation from a time-dependent Schr\"odinger equation, which can be used to calculate the nonlinear vibrational response \cite{Torii.2006.jpca.110.4822, Jansen.2009.acr.42.1405}, this approach does not describe the quantum mechanical nature of the bath. The approximation of a fast bath (often confusingly referred to as Markovian), commonly used in the case of electronic excitations neglects, as mentioned before, the bath memory, and therefore does not describe the partially inhomogeneous broadening observed in experiments, as has been shown in the case of a single spin system \cite{Ishizaki.2008.chemphys.347.185}. While the memory can be correctly included in the case of mostly static fluctuations \cite{Dijkstra.2008.jcp.128.164511}, we shall develop here the full calculation for multiple coupled two-level systems in the presence of a bath which evolves on an arbitrary time scale.  

The third-order response can be measured in most detail by exciting the system with three short laser pulses, which interact at times $\tau_1$, $\tau_2$ and $\tau_3$. In-between the pulses, the time evolution is dictated by the Hamiltonian $H_\mathrm{M}$. We define variables for the evolution times by $t_1 = \tau_2 - \tau_1$, $t_2 = \tau_3 - \tau_2$ and $t_3 = t - \tau_3$. In the same way as in the linear response, the use of short excitation pulses allows the measurement of the third-order response function, which is given by
\begin{equation}
  S^{(3)}(t_1, t_2, t_3) = i^3 \langle \tr \mu G(t; \tau_3) \xop \mu G(\tau_3; \tau_2) \xop \mu G(\tau_2; \tau_1) \xop \mu R(t_0) \rangle_O.
\end{equation}
Because the sample size is typically much larger than the wavelength of the light, the signal emitted in a third-order experiment is found in specific directions, given by linear combinations of the wavevectors of the incident pulses \cite{Mukamel.2000.arpc.51.691}. This allows the selection of a part of the response function. As mentioned before, in the impulsive excitation limit, it is necessary to apply the RWA for the system laser interaction. In the photon echo geometry, where the wave vector of the signal is equal to the sum of the wave vectors of pulse two and three, minus the wave vector of pulse one, three contributions survive the RWA. The response function is the sum of these terms, which can be interpreted as a ground state bleach process, a stimulated emission process, and an induced absorption process. Their contributions to the response functions are given by
\begin{eqnarray}
   S_\mathrm{GB}(t_1, t_2, t_3) &=& i^3 \langle \vec \mu_{m'} \vec \mu_m G_{m'0; m0}(t; \tau_3)  \sigma \rangle_O, \nonumber \\
   S_\mathrm{SE}(t_1, t_2, t_3) &=& i^3 \langle \vec \mu_{m'''} \vec \mu_{n''} G_{m'''0;m''0}(t; \tau_3) \rho_{m''n''} \rangle_O, \nonumber \\
   S_\mathrm{IA}(t_1, t_2, t_3) &=& - i^3 \langle \vec \mu_{w'n'''} \vec \mu_{wm''} G_{w'n''';wn''}(t; \tau_3) \rho_{m''n''} \rangle_O,
   \label{nlresponse}
\end{eqnarray}
respectively. 
where we have defined
\begin{equation}
  \sigma = \vec \mu_{n'} \vec \mu_n G_{0n';0n}(\tau_2; \tau_1)
\end{equation}
and
\begin{equation}
  \rho_{m''n''} = \vec \mu_{m'} \vec \mu_n G_{m''n'';m'n'}(\tau_3; \tau_2) G_{0n';0n}(\tau_2; \tau_1).
\end{equation}
In these equations, repeated indices are understood to be summed over. The orientational average over the rank four tensor formed by the product of four dipoles can be written in Cartesian components as $\langle \vec \mu \vec \mu \vec \mu \vec \mu \rangle_O = \sum_{\alpha\beta\gamma\delta} A_{\alpha\beta\gamma\delta} \mu^\delta \mu^\gamma \mu^\beta \mu^\alpha$. The tensor $A$ contains the effects of polarization of the laser pulses. Three fundamentally different combinations of polarizations can be used, giving different values for the elements in $A$ \cite{Hochstrasser.2001.chemphys.266.273}. Combined with the sum over Cartesian components, labelled with $\alpha, \beta, \gamma, \delta$, they describe the averaging of the signal over an isotropic sample. 

During the coherence time $t_3$, the induced absorption contribution contains coherent superpositions between one- and two-exciton states. These two-exciton states are labelled with indices $w$ and $w'$ in the last line in (\ref{nlresponse}).
We work in the site basis, where the two-exciton basis states are given as the direct product of two excitations. For example, a two-exciton state can be written as $|w\rangle = |w_1 w_2\rangle$, in which case base $w_1$ and $w_2$ are excited, while the electrons in all other bases are in the ground state. Consequently, the transition dipoles between one- and two-exciton states are given by $\vec \mu_{wn} = \vec \mu_{w1w2,n} = \delta_{w1,n} \vec \mu_{w2} + \delta_{w2, n} \vec \mu_{w1}$. 

It is important to note that the above-mentioned memory stored in the bath is fully included in these expressions for the nonlinear response. The propagators $G$ are matrices in the product basis of system and bath degrees of freedom, and multiplications over the bath degrees of freedom are understood by writing the explicit dependence of the propagators on two times. In a more complete but complicated notation, one might write for example
\begin{equation}
  \rho_{m''n''}^{U, \alpha\beta} = \mu_{m'}^\beta \mu_n^\alpha G^{U;T}_{m''n'';m'n'}(t_2) G^{T;S}_{0n';0n}(t_1) R^S_0,
\end{equation}
where $S, T$ and $U$ denote basis states in the Liouville space of the bath degrees of freedom. $S$ and $T$ are understood to be summed. This dependence of the propagators, which describes the memory of the bath and of the system-bath coherence, is correctly included by propagating the hierarchy of equations of motion. The simulation procedure of, for example, the stimulated emission contribution, is as follows. First, the matrix elements of the reduced density matrix, as well as all the auxiliary density matrices, between a singly excited state and the ground state, are populated according to their transition dipole. The hierarchy of equations of motion is then solved for a time $t_1$. The second multiplication with the transition dipoles gives matrix elements of all the (auxiliary as well as physical) density matrices between two singly excited states. The hierarchy for these matrices is propagated for a time $t_2$. A third multiplication with the dipoles again gives a coherence between singly excited states and the ground state, which is propagated for a time $t_3$, followed by a final transition dipole multiplication and a trace operation to obtain the signal. This procedure is repeated for all required combinations of Cartesian components.

While the linear absorption probes only the dynamics of coherent superpositions between the ground state and one-exciton states, the third-order response function contains the dynamics of matrix elements $\langle n | \rho | m \rangle$ during $t_2$. These terms describe the motion of a single excitation in the system, as well as the time evolution of quantum coherences between system states. The excitation dynamics can therefore be measured as a function of $t_2$, and be correlated with the time evolution of the coherences during $t_1$ and $t_3$.

\subsubsection{Two-time anisotropy}
A useful measurement of this information in a system where the transition dipoles of individual bases point in various directions, as is the case in DNA, is the two-time anisotropy decay. It can be found by setting $t_3 = 0$. In this case, the stimulated emission is cancelled by a contribution from the induced absorption, and the signal simplifies considerably,
\begin{equation}
  S(t_1, t_2) = \left\langle \vec\mu_m \vec\mu_m \sigma - \sum_{p \neq n} \vec\mu_p \vec\mu_p \rho_{nn} \right\rangle_O.
\end{equation}
This anisotropy can be measured independently in the parallel polarization geometry (using four pulses all polarized in the same direction), which gives a signal $S^\mathrm{par}$, and in the perpendicular geometry (where the final pulse pair is polarized perpendicularly to the first pair), which yields $S^\mathrm{per}$. In simuations, these two reponse functions are found by choosing the appropriate coefficients $A$. The two-time anisotropy is then calculated as
\begin{equation}
  S_\mathrm{TTAD}(t_1, t_2) = \frac{S^\mathrm{par}(t_1, t_2) - S^\mathrm{per}(t_1, t_2)}{S^\mathrm{par}(t_1, t_2) + 2 S^\mathrm{per}(t_1, t_2)}.
\end{equation}
The anisotropy describes the average rotation of the dipole of the excitation during the times $t_1$ and $t_2$. It is a complex valued quantity, and in the rest of this paper we will only consider its absolute value.
In numerical calculations, it is an advantage that the propagation of the two-exciton states is not required.

\subsection{Static limit}
The time scale of the bath $\tau_B$ is modelled by the parameter $\gamma=1/\tau_B$. The static limit is recovered by assuming an ultra slow bath, characterized by the limit $\gamma \to 0$. The correlation function becomes
\begin{equation} \label{cfstat}
 L(t) = \frac{2 \lambda}{\beta}.
\end{equation}
We see that the imaginary part of the correlation function vanishes in this limit, while the real part is a constant. In physical terms, this limit means that fluctuations in the base transition energies are now frozen. The system can be viewed as an ensemble of helices, in each of which the bases have different energies. In addition, in the case of incomplete correlation, each base within a given helix has a different transition frequency. The system is therefore no longer ergodic, as should be expected in the static limit. The absence of an imaginary term in the correlation function shows that there is no dissipation. This can be explained by considering that the bath modes are now ultra slow; they have too low frequencies to accept any energy from the system in a finite time.

In the static limit, the linear absorption spectrum can be calculated from Fermi's golden rule,
\begin{equation}
  A(\omega) = \frac{1}{3} \langle \sum_k f_k \delta(\omega - E_k) \rangle, 
\end{equation}
where $E_k$ is the energy of the $k^\mathrm{th}$ eigenstate with eigenfunction $\phi_k$, and $f_k = \left( \sum_{n} \phi_{kn} \vec \mu_n\right)^2 $ is its oscillator strength. The average, indicated by the notation $\langle \cdots \rangle$ is over realizations of the static disorder. Similar expressions can be derived for the non-linear response.

The dynamics of a single excitation in the site basis can be found by projecting the initial condition on the eigenstates, propagating the density matrix in the eigenbasis and projecting back. In each realization, an initial population on site $n$ gives a density matrix $\rho_{kq}(0) = \phi_{kn} \phi_{qn}$. Its time evolution is given by
$\rho_{kq}(t) = \exp(-i(E_k - E_q)t) \rho_{kq}(0)$, and the final density matrix in the site basis is obtained as $\rho_{nm}(t) = \sum_{kq} \phi_{kn} \phi_{qm} \rho_{kq}(t)$. This density matrix can be averaged over realizations to obtain the average population on a given site, or the coherence between sites.

\section{Results and Discussion} \label{sec:results}
\subsection{Delocalization}
The properties of elementary excitations in the DNA are determined by a competition between electronic interactions between bases, which favour extended states, and environment-induced fluctuations, which lead to localization. In the case of DNA, the coupling to the environment, which is determined by the reorganization energy and the temperature according to (\ref{cfstat}), is an order of magnitude larger than the electronic couplings. We therefore analyse our results in the site basis, and discuss populations on individual bases and coherences between bases. One might wonder if the large reorganization energy does not prohibit any significant quantum coherence between bases. 
In the static limit, the properties of the system can be calculated by averaging over all members in the ensemble, or, equivalently, over individual realizations of the disorder. For each realization, the Hamiltonian can be diagonalized to yield the eigenstates. These eigenstates are used to calculate the observable of interest, which is finally averaged over the realizations of the disorder. The localization length in a linear chain can be estimated using the properties of the eigenstates \cite{Malyshev.1993.jlumin.55.225, Malyshev.2001.prb.63.195111}. Such an estimate would lead to the conclusion that excitations are spread over a few bases. Here, we directly calculate the coherence between two bases in an ensemble of dimers, each consisting of only two neighbouring A (or, equivalently, T) bases. For a particular system in the ensemble, the Hamiltonian can be written as $H = (\delta\epsilon/2)(\+{c}_2 c_2 - \+{c}_1 c_1) + J (\+{c}_1 c_2 + \+{c}_2 c_1)$. 

While the value of the coupling $J$ is the same throughout the ensemble, the energy difference between the two bases $\delta\epsilon$ varies from one system to another. This difference, which originates in the fluctuations, is characterized with a Gaussian distribution, which has a standard deviation $\sigma = \sqrt{2 \lambda / \beta}$. In each realization, diagonalization of the Hamiltonian gives the eigenstates $|+\rangle$ and $|-\rangle$ (note that these are the symmetric and anti-symmetric states only in the absence of disorder). For a positive value of $\delta\epsilon$, the state with the highest energy is found mostly on base 2, while for negative $\delta\epsilon$ it occupies mainly base 1. In either case, the coherence $C = \langle 1|+\rangle \langle +|2\rangle$ is well-defined, and given by
\begin{equation}
  C = \frac{J(-\delta\epsilon + \sqrt{D})}{D - \delta\epsilon \sqrt{D}},
\end{equation}
with the energy gap between the eigenstates $D = \sqrt{\delta\epsilon^2 + 4 J^2}$. Integrating over a Gaussian distribution, we find the average coherence in the presence of disorder to be
\begin{equation}
  \langle C \rangle = \frac{J/\sigma}{\sqrt{2 \pi}} e^{(J/\sigma)^2} K_0((J/\sigma)^2),
\end{equation}
where $K_0$ is a modified Bessel function of the second kind. In the absence of disorder, which amounts to taking the limit of $\sigma/J$ to zero, the coherence reduces to the value for completely delocalized states, $\langle C \rangle = 0.5$. In the other limit, where $\sigma / J \to \infty$, complete localization destroys the coherence, which approaches zero. Surprisingly, however, substantial coherence is present for large values of the disorder. In particular, for $\sigma/J = 10$, we find $\langle C \rangle = 0.19$. This shows that, even though the coupling to the environment is quite strong compared to the electronic interactions between the bases, the excitations cannot be understood as completely localized. Even though, based on the estimate mentioned before as well as on earlier 
work \cite{Bouvier.2003.jpcb.107.13512}, the localization length is not expected to be larger than a few bases, the coherence between bases cannot be ignored in a description of energy transport.

\subsection{Linear absorption spectrum}

\begin{figure}[t]
 \includegraphics{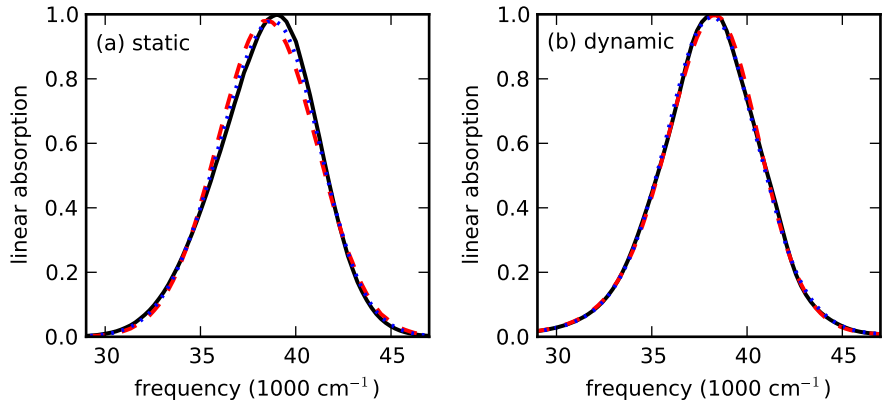}
\caption{\label{fig:lin} Linear absorption spectra $A(\omega)$ in (a) the static limit and (b) with dynamic fluctuations for uncorrelated baths (solid line), correlated fluctuations in each base pair (dotted line) and fully correlated fluctuations (dashed line).}
\end{figure}

The linear absorption spectrum, calculated from (\ref{linspec}), is shown in figure \ref{fig:lin} for the static as well as the dynamic environment. The static results were calculated by averaging over $10^6$ disorder realizations. We observe that the line shape appears similar in all three models, but that the correlations lead to a shift in the absorption maximum in the case of a static environment. This can be understood from the decrease of the localization in the presence of correlations. For completely correlated disorder, the eigenstates are delocalized over all available bases. The delocalization changes the character of the bright states, which leads to a shift in the absorption spectrum. 

The spectra for a dynamic bath are shown in panel (b). The spectra for correlated and uncorrelated fluctuations appear almost completely similar; the shifts observed in the static spectrum disappear. We conclude that the presence of correlation cannot conclusively be established from the linear absorption spectrum.

\subsection{Exciton dynamics}

\begin{figure}[t]
 \includegraphics{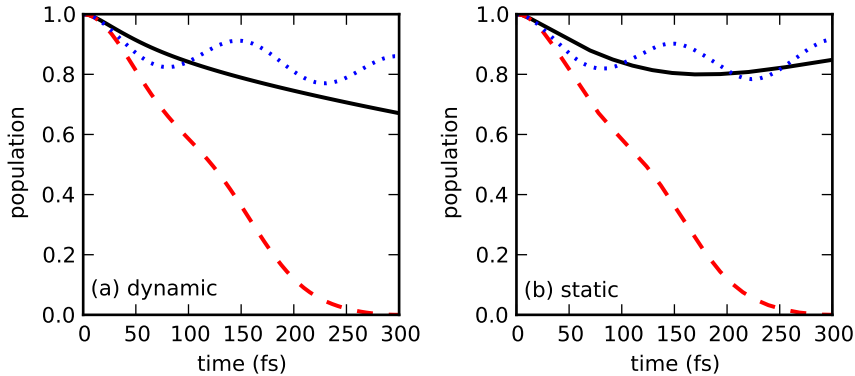}
\caption{\label{fig:population} Population on the first A base, $\rho_{1A1A}(t)$, after initial excitation of this base for uncorrelated fluctuations (solid line), correlated fluctuations in each base pair (dotted line) and completely correlated fluctuations (dashed line).}
\end{figure}

Although the difference between correlated and uncorrelated fluctuations can't be seen in the linear absorption, it has a strong effect on the exciton dynamics. This can be clarified by performing a numerical experiment, in which the first A base is artificially excited. The subsequent time dependence of the population on this site is shown in figure \ref{fig:population}.
The fully correlated case is rather trivial. In this case, the fluctuations have no effect at all, and the dynamics are generated by the system's Hamiltonian without influence from the bath. The time scale of the bath is consequently not important, the time evolution is the same for all values of $\gamma$. For perfectly correlated fluctuations in the A and T base in each base pair, but no correlations between the base pairs, the exciton dynamics is, perhaps surprisingly, slower than for completely uncorrelated fluctuations. This effect is only present in the dynamical model for the environment. As expected, the dynamic and static baths give rise to almost the same exciton dynamics in the first 100 fs, while significant differences appear for longer times. This is especially clear in the population dynamics in the case of uncorrelated fluctuations. 

\begin{figure}[t]
 \includegraphics{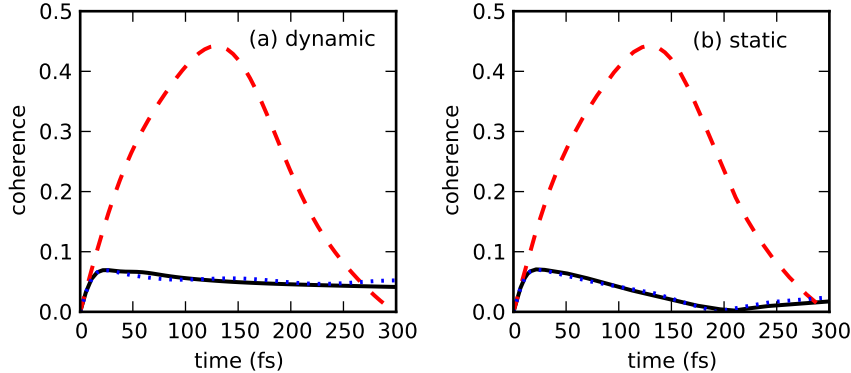}
\caption{\label{fig:coherence} Absolute value of the coherence between neighbouring A bases, $\rho_{1A2A}(t)$, after initial excitation of the first A for uncorrelated fluctuations (solid line), correlated fluctuations in each base pair (dotted line) and completely correlated fluctuations (dashed line).}
\end{figure}

The coherence between two neighbouring A bases is plotted in figure \ref{fig:coherence}. The coherence is clearly increased by the presence of correlations along the stack. The real part of the coherence (not shown) is much larger in the presence of dynamic fluctuations as compared to a static bath, when it is zero for a dimer. We find similar results for the coherence between the A and T bases (not shown).

\subsection{Two-time anisotropy decay}
The time dependence of the anisotropy, as measured in a third-order non-linear experiment, contains information about the exciton dynamics, and has been used for that reason to study electronic excitations in conjugated polymers \cite{Collini.2009.science.323.369}, as well as vibrational modes in peptides \cite{Jansen.2006.jpcb.110.22910}.
The presence of disorder has been shown to lead to relaxation of the anisotropy within 100 fs in conjugated polymers \cite{Dykstra.2009.jpcb.113.656}. This effect could be explained by interaction with an environment which contains either a slow (static) or a fast mode, compared to the time scale of the system's dynamics as probed in the experiment. In the presence of static disorder, interaction with light creates excitation of eigenstates of the system, as well as coherent superpositions of these eigenstates. The populations in the eigenbasis do not evolve under the system Hamiltonian, while the coherences $\rho_{kq}$ only collect a phase factor determined by the energy difference $E_k - E_q$. Incoherent relaxation is introduced by the fast environmental modes, which are treated perturbatively. Because of the assumption that these modes relax much faster than any time scale in the system, a master equation can be derived for the populations of each eigenstate. Such a description has been successfully applied to understand the dynamics of excitons in self-assembled aggregates at a temperature low enough to freeze the solvent, in which case the assumption of a static environment is valid \cite{Heijs.2005.prl.95.177402}.

If this type of calculation is applied to our current results of the anisotropy relaxation in the first 100 fs, this assumption would imply that the fast environmental modes have characteristic dynamics on time scales considerably smaller than 10 fs. 
Here, we rather model the decay of the anisotropy using a single time scale $\tau_B$ of the bath. For times smaller than $\tau_B$, a static description of the dynamics is valid. The dynamics are then governed only by coherent evolution under the system Hamiltonian in each static realization. For times comparable to or larger than $\tau_B$, the bath time scale leads to the destruction of the initial eigenstates, and the results will deviate from a picture with static disorder only. 

\begin{figure}[t]
 \includegraphics{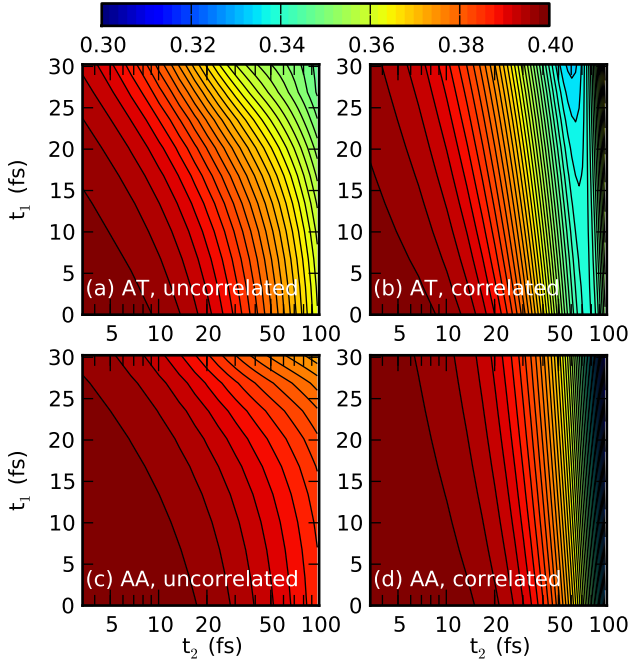}
\caption{\label{fig:anisotropysingle} Two-time anisotropy decay $S_\mathrm{TTAD}(t_1, t_2)$ for (a) a single base pair with uncorrelated fluctuations (b) a single base pair with correlated fluctuations, (c) two neighbouring A bases with uncorrelated fluctuations and (d) two neigbouring A bases with correlated fluctuations.}
\end{figure}

\begin{figure}[t]
 \includegraphics{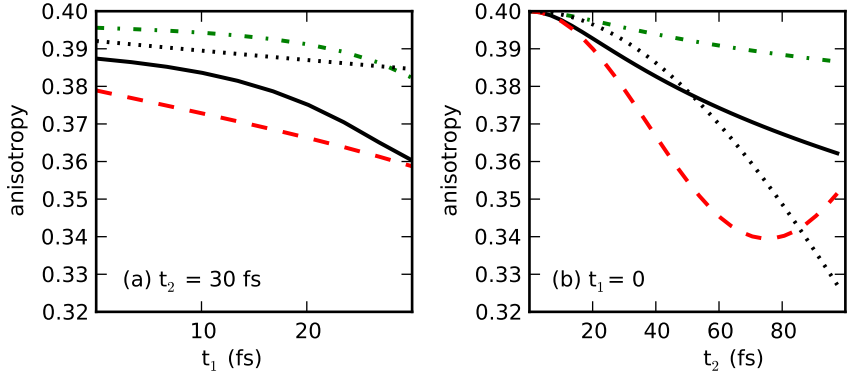}
\caption{\label{fig:anisotropysingleslices} Slices through the two-time anisotropy for a single base pair with uncorrelated fluctuations (solid line) and correlated fluctuations (dashed line) as well as for two neighbouring A bases with uncorrelated fluctuations (dash-dotted line) and correlated fluctuations (dotted line).}
\end{figure}

Because our model does not include the (presumably slow) rotation of the DNA molecule, the anisotropy does not decay for a single base. All the decay in the anisotropy can therefore be ascribed to energy transport in the DNA. The anisotropy decay in the DNA model is quite complex. It involves transfer between the A and T bases in a pair, which leads to a large (117 degrees) rotation of the transition dipole, as well as exciton migration along the stack, which is energetically more favourable, but only rotates the transition dipole by 36 degrees in a single hop. It is not a priori clear which mechanism will contribute more to the anisotropy decay. To understand these two contributions separately, we present the decay of the anisotropy in a single base pair, as well as in two neighbouring A bases in figure \ref{fig:anisotropysingle}. In general, the anisotropy decays faster for transfer between the A and T base, for both correlated as well as uncorrelated fluctuations. This indicates that in DNA, the effect of the larger dipole rotation is more important than the difference in the A and T excitation energies. In the case of fully correlated fluctuations, however, the recurrence time in the AT base pair is shorter than in the AA stacked pair, as can be seen in the cross sections in figure \ref{fig:anisotropysingleslices}. Furthermore, the presence of fluctuations always slows down the initial decay of the anisotropy, as can be seen from the faster decay in the case of correlated as compared to uncorrelated fluctuations.

\begin{figure}[t]
 \includegraphics{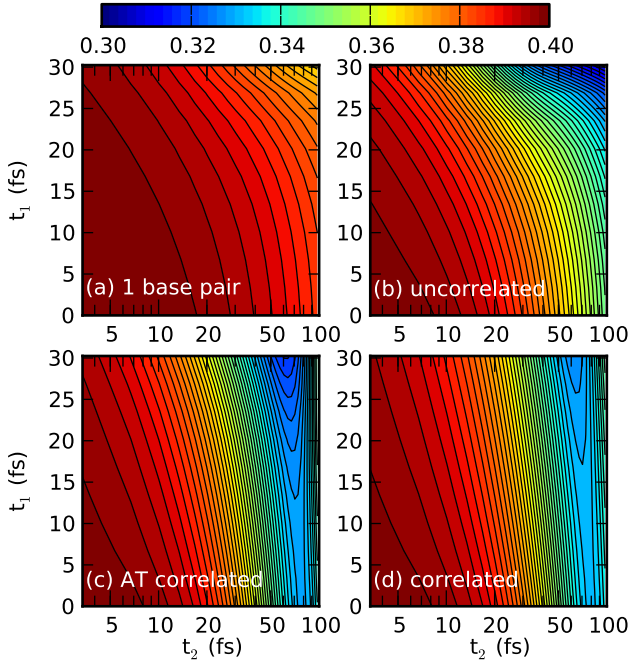}
\caption{\label{fig:anisotropy} Two-time anisotropy decay $S_\mathrm{TTAD}(t_1,t_2)$ for (a) a single base pair, (b-d) 12 bases with (b) uncorrelated fluctuations in each base, (c) fully correlated fluctuations in each base pair, but no correlations between base pairs and (d) fully correlated fluctuations.}
\end{figure}

We now turn to the discussion of the two-time anisotropy decay in DNA, calculated with a bath time scale of 50 fs. Comparing the results with the decay of the anisotropy in a single base pair with uncorrelated fluctuations (for clarity repeated in figure \ref{fig:anisotropy} (a)), we see that the decay is faster for the extended helix. The excitation now has the freedom to move along the stack as well as between base pairs, which explains the faster decay. We next analyse the effect of correlations in the fluctuations, which clearly leads to differences in the two-time anisotropy results.

\begin{figure}[t]
 \includegraphics{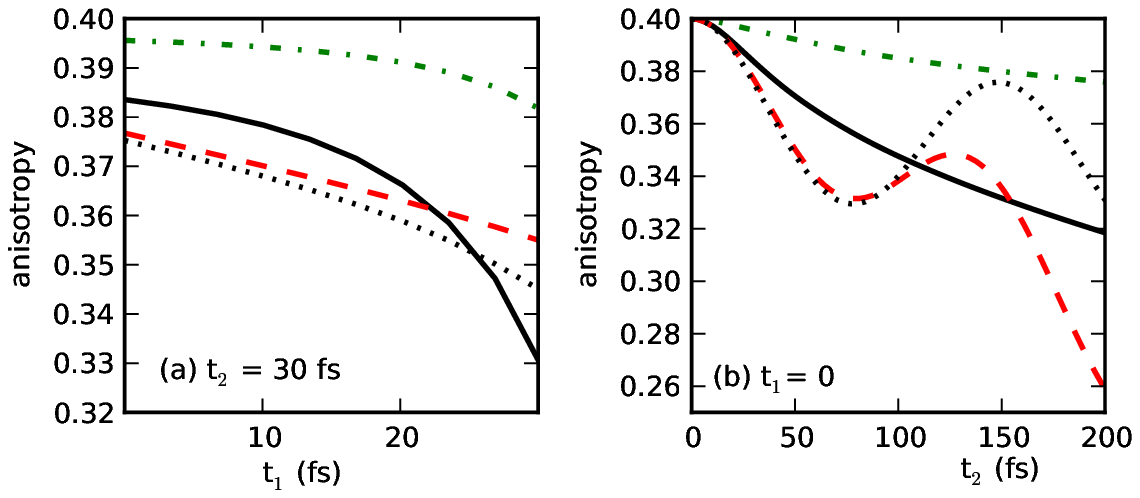}
\caption{\label{fig:anisotropyslices} Slices through the two-time anisotropy for a single base pair (dash-dotted line) and a helix with 12 bases and uncorrelated fluctuations in each base (solid line), correlated fluctuations within each base pair (dotted line) and fully correlated fluctuations (dashed line).}
\end{figure}

For $t_1 = 0$, plotted in figure \ref{fig:anisotropyslices} (b), the fully correlated and the AT correlated cases show similar results. In this case, the transfer within a base pair is completely coherent, as can be seen from the presence of a minimum near $t_2 = 80$ fs, after which the anisotropy increases again. This is the same coherent recurrence as observed in the populations and coherences before. In the presence of uncorrelated fluctuations, such recurrences are not present. In this case, the anisotropy decays steadily, in line with the behaviour of the population in a dynamic bath shown in figure \ref{fig:population} (a). We conclude that the anisotropy measurement contains clear information on the presence of correlated fluctuations.

Because our simulations treat only a small piece of DNA, one might expect finite size effects to play a significant role. To address this, we have repeated our simulations for a helix with five base pairs, and found almost identical results. We therefore conclude that finite size effects do not significantly change our conclusion and believe that our study of the short time dynamics provides insight in the properties of real DNA samples.

\section{Concluding Remarks} \label{sec:concl}

In conclusion, we have modelled the exciton dynamics and non-linear optical response in a model BDNA helix, in the presence of a dynamic bath. In this study, we have used the hierarchy of equations of motion approach, which nonperturbatively includes the environment-induced fluctuation and dissipation, and extended it for the first time to include correlations in the fluctuations. The correlations have only a small effect on the static linear absorption spectrum, and are not visible in the linear absorption in the presence of a dynamic environment. We find that the presence of correlation in the fluctuations does have a strong effect on the exciton dynamics, which is reflected in the two-time anisotropy decay.

Our current simulations are limited to a small piece of the DNA helix. We have found that the system size does not significantly influence our results. However, for the treatment of other systems, an extension of the method presented here to a larger system size is desirable. Although this is in principle straightforward, the simulation of larger systems is computationally expensive. 
In the case of DNA, the high temperature approximation is applicable, which speeds up the computation. However, because of the large ratio of the reorganization energy to the coherent coupling, the number of tiers in the hierarchy must be rather large, especially if we want to extend the calculation to a slower bath. The deeper tiers include terms which involve the action of multiple system-bath operators on bases that are separated in space. Correlations between the excitations on bases which are separated by more than the localization length are expected to be small.
Computational improvements based on this physical argument, be it automatic truncation of the hierarchy as proposed recently \cite{Shi.2009.jcp.130.084105}, or new algorithms that take into account the localization of exciton states explicitly, are desirable.
Furthermore, charge transfer states may be important in DNA \cite{Bittner.2006.jcp.125.094909, Conwell.2008.jpcb.112.2268}, and could be included in further development of the calculations presented here.

The helical structure is important for a correct description of the anisotropy decay. For a precise comparison with experimental results, it is therefore desirable to include variations in the structure. Finally, while this work was limited to poly(dA)poly(dT) DNA, a straightforward extension of the method to include other base sequences would help in the interpretation of experiments.

\end{document}